%% file: hdf.tex
\documentstyle{mn}
\input {epsf_mn.tex}

\def\figinsert#1#2{\epsfbox{#1} \message{#2} }    
\begin{document}
\title[AGN predictions for the Hubble Deep Field and the X-ray background]
{AGN predictions for the Hubble Deep Field and the X-ray background}
\author[O. Almaini and A.C. Fabian ]
{O.~Almaini and A.C. ~Fabian\\
Institute of Astronomy, Madingley Road, Cambridge, CB3 OHA}

\date{MNRAS in press}
\maketitle

\begin{abstract}
We estimate the number of AGN among the galaxies in the Hubble Deep
Field (HDF). The recent discovery of a class of X-ray luminous narrow
emission-line galaxies has provided a possible explanation for the
origin of the X-ray background (XRB), although the nature of the
activity is still unresolved. By extrapolating the observed X-ray
number count distribution to faint optical magnitudes we predict that
this AGN population could account for a significant fraction, $\sim
10$ per cent, of the galaxies in the HDF. In contrast, normal broad
line QSOs are expected to account for no more than $\sim 0.1 $ per
cent of the sources at these magnitudes.

\end{abstract}


\section{Introduction}

Little is known about low luminosity or obscured (i.e. Seyfert 2) AGN
other than in the very local universe. Deep optical observations such
as the HDF (Williams et al. 1996) provide an opportunity to probe such
objects at moderate to high redshift.  Locally, a few percent of
galaxies have active nuclei and if this holds true at $z \sim 2$ then
there could be several tens of AGN in the HDF.

Low luminosity and/or obscured AGN are also prime candidates for
explaining the origin of the XRB. In particular, it is becoming clear
from deep $\em ROSAT$ observations that a population of X-ray luminous
narrow emission-line galaxies (NELGs) could dominate the source counts
at faint X-ray fluxes (Boyle et al. 1995a, Roche et al. 1995, McHardy
et al. 1995, Carballo et al. 1995, Griffiths et al. 1996). Following
Hasinger (1996) we will refer to these X-ray loud objects as
`narrow-line X-ray galaxies' (NLXGs) in order to distinguish them from
the other narrow emission-line galaxies found in increasing numbers in
deep optical surveys (eg. Tresse et al. 1996). The discovery of strong
evolution in this new X-ray population (Boyle et al. 1995a, Griffiths
et al. 1996, Almaini et al. 1997) can now allow us to make predictions
for their contribution at faint X-ray fluxes and hence extrapolate to
deep optical surveys.  We make such predictions and find that NLXGs
could outnumber ordinary broad line QSOs in the HDF by approximately
two orders of magnitude and plausibly account for $\sim 10$ per cent
of the galaxies in the HDF. We suggest that workers on the HDF should
be aware of this possibility in accounting for any unusual properties
of these faint galaxies. We assume $q_0=0.5$ and $H_o=50 $km
s$^{-1}$Mpc$^{-1}$ throughout.

\section{The X-ray source source populations}

The origin of the X-ray background is still a major unsolved problem,
but at soft energies ($< 3.5$ keV) great strides have been made
following the launch of the Einstein and ROSAT satellites. By
resolving as many sources as possible in the deepest $\em ROSAT$
exposures, up to $70$ per cent of the XRB can now be resolved into
discrete sources.  By optical spectroscopy at least half of the
detected sources have been identified as broad-line QSOs, which
directly account for $\sim 30 $ per cent of the XRB at 1keV (Shanks et
al 1991).

As larger samples of QSOs became established, studies of the X-ray
luminosity function (Boyle et al. 1994) and the source number count
distribution (Georgantopoulos et al. 1996a) have shown that QSOs are
unlikely to contribute more than $50$ per cent of the XRB at
1keV. This suggested the existence of another source population. A
further problem in explaining the XRB with QSOs is the shape of their
X-ray spectra. QSOs show relatively steep X-ray spectra with indices
of $\Gamma=2.2\pm0.1$ while the $1-10$\,keV XRB has a significantly
flatter spectrum with $\Gamma=1.4$ (Gendreau et al. 1995).  Recent deep
$\em ASCA$ GIS observations in the harder $2-10\,$keV X-ray band have
found a number count distribution roughly a factor of two above the
$\em ROSAT$ source counts (Georgantopoulos et al. 1996b, Inoue et al
1996). This also suggests the presence of another population at hard
energies, other than the broad-line AGN which dominate the bright $\em
ROSAT$ counts.

The number count distribution of X-ray sources in the $\em ROSAT$ band
can be well fitted by a Euclidean power law ($\gamma=2.5$) at bright
X-ray fluxes but turns over to a flatter power law slope of $\gamma
\sim 1.5$ below $S(0.5-2.0$\,keV$) \simeq2 \times 10^{-14}
$erg$\,$s$^{-1}$cm$^{-2}$ (Hasinger et al. 1993, Branduardi-Raymont et
al. 1994, Vikhlinin et al. 1995).  Separating QSOs from this
population, their fractional contribution appears to fall below this
limit as they turn over to an even flatter power law slope of $\gamma
\sim 1$ (Georgantopoulos et al. 1996a). This is confirmed by a detailed
analysis of the X-ray luminosity function (Boyle et al. 1993, 1994)
where the total QSO contribution was found to lie in the range $34-53$
per cent, with the main uncertainty depending on the faint end slope.

There is now very strong evidence that the `new' X-ray source
population consists of unusually X-ray luminous galaxies, typically
$10-100$ times brighter than normal field galaxies, which could
account for the remainder of the XRB (Boyle et al. 1995a, Roche et al
1995, Carballo et al. 1995, Griffiths et al. 1996, McHardy et al
1997). The discovery that these galaxies have hard X-ray spectra
(Carballo et al. 1995, Almaini et al. 1996, Romero-Colmenero et
al. 1996) lends further weight to hypothesis.

Preliminary estimates of luminosity functions at the bright end of the
NLXG population have shown evidence for strong evolution,
parameterized by the form $L_X \propto (1+z)^3$ (Boyle et al
1995b, Griffiths et al. 1996).  Evidence for strong evolution in the
fainter population has recently been obtained by Almaini et al. (1997)
by cross-correlating the unresolved XRB in deep $\em ROSAT$ exposures
with faint galaxy catalogues.  By using the optical magnitude of the
catalogue galaxies as probes of their redshift distribution, evidence
for strong evolution was found of the form $L_X \propto
(1+z)^{3.2\pm1}$ (Almaini et al. 1997).

\begin{figure}
\centering \centerline{\epsfxsize=12truecm \figinsert{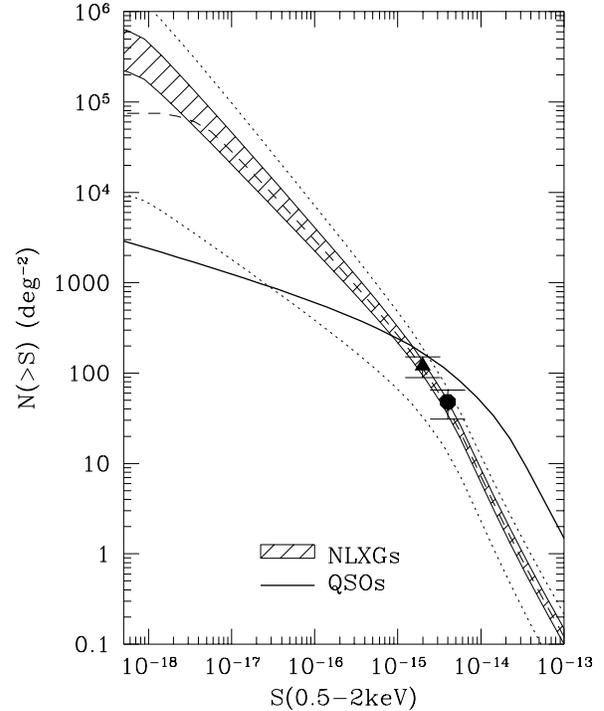}{0.0pt}}
\caption{Showing the predicted number-flux relation for broad line
QSOs and NLXGs. QSO predictions are based on the luminosity functions
of Boyle et al. (1994).  The dotted lines reflect the $1\sigma$ errors
on the faint end slope of the NLXG luminosity function of Griffiths et
al (1996).  The shaded area shows the tighter predictions which arise
if we impose the fainter measurements by Almaini et al. (1997) (filled
circle) and Hasinger (1996) (filled triangle).  The dashed line shows
the effect of changing the lower limit in the luminosity function from
$10^{39}$erg s$^{-1}$ to $10^{40}$erg s$^{-1}$.}
\end{figure}

We will now  estimate the likely contribution
of X-ray luminous galaxies to the HDF. We use the faintest luminosity
function available to date, by Griffiths et al. (1996), who collated a
sample of 32 NLXGs from both $\em ROSAT$ and $\em Einstein$
surveys. They find a best fitting local ($z=0$) luminosity function
modelled by a broken power law of the form:

\begin{equation}
\Phi_x(L_x)=\left\{\begin{array}{ll}
K_1L_x^{-\gamma_1} & L_x(z=0)<L_x^* \\
K_2L_x^{-\gamma_2} & L_x(z=0)>L_x^* \\
\end{array}
\right.
\end{equation}

\noindent where  $K_1$, $\gamma_1$ and $\gamma_2$ represent the
normalization, faint and bright end slopes of the XLF
respectively.  The best fitting values were found to be
$\gamma_1=1.85\pm0.25$, $\gamma_2=3.83\pm0.20$ and
$K_1=1.3\times10^{-7}$Mpc$^{-3}(10^{44}$erg$^{-1})$. The `break'
luminosity, $L_x^*$, was found to evolve with the form:

\begin{equation}
L_x^* \propto (1+z)^{3.35\pm0.25}.
\end{equation}

This is in good agreement with the evolution found in fainter NLXGs by
Almaini et al. (1997). We use this luminosity function to calculate
the predicted NLXG number-flux relation to very faint limits.  We
integrate this model out to $z=4$, with no evolution beyond
$z_{max}=2$, as observed for broad-line AGN (eg. Boyle et al 1994).
We use a local luminosity function defined over the range
$10^{39}-10^{45}$erg s$^{-1}$. We ignore NLXGs with X-ray luminosities
below $10^{39}$erg s$^{-1}$ since these appear to have more `normal'
X-ray to optical ratios (eg. Dela-Ceca et al. 1996) and are therefore
less likely to be AGN.  A correction was made to convert to the
$0.5-2\,$keV $\em ROSAT$ band by assuming an X-ray spectral index of
$\Gamma=1.4$, as required if these objects are to explain the XRB.
This also modifies the evolution to the form $L_x^* \propto
(1+z)^{2.75\pm0.25}$. 

The resulting predicted faint source counts are displayed in Figure 1.
The dominant uncertainty in the extrapolation is the slope of the
faint end of the luminosity function. The dotted lines in Figure 1
show the effect of varying this parameter within its $1\sigma$ error.
Integrating this distribution to the faintest limits ($10^{-20}
$erg$\,$s$^{-1}$cm$^{-2}$) we obtain an NLXG contribution of $\sim
25^{+\infty}_{-13}$ of the XRB at 1keV (the upper limit formally
diverges and saturates the residual XRB at $\sim 10^{-17}
$erg$\,$s$^{-1}$cm$^{-2}$).  Throughout this paper we assume that the
intensity of the XRB at 1keV is $12.0$ keV
cm$^{-2}$s$^{-1}$sr$^{-1}$keV$^{-1}$ (Georgantopoulos et al. 1996a).  The
solid line represents the predicted number-flux relation for
broad-line QSOs, based on Model U in Boyle et al. (1994). This model
predicts that QSOs account for $\sim 40 \pm 10$ per cent of the XRB at
1keV, although for clarity we have not plotted the uncertainties in
the expected source counts.

It is clear that extrapolating the Griffiths et al.  luminosity
function gives very uncertain faint source count predictions.  Fainter
observations are required.  In addition, as described in Griffiths et
al. (1996), a possible source of error is the incompleteness of the
$\em ROSAT$ survey which defines the faint end of their sample. In
many cases source confusion hindered the unambiguous identification of
the faintest X-ray sources and so a first order correction was applied
following Boyle et al. (1993).  Recent measurements of the number of
NLXGs at fainter limits than those probed by Griffiths et al. can
allow us to constrain the predicted number-flux relation more tightly.
Almaini et al. (1997) used a statistical method to overcome confusion
problems directly. By cross-correlating the X-ray sources not firmly
identified with QSOs or stars with faint optical galaxy catalogues
they obtained a significant positive signal, the amplitude of which
implies a surface density of $48\pm17$\, X-ray galaxies deg$^{-2}$ at
a flux of $4\times 10^{-15} $erg$\,$s$^{-1}$cm$^{-2}$.  With the
improved resolution of the $\em ROSAT$ HRI, the confusion problems are
considerably reduced. Recent ultra-deep HRI observations by Hasinger
(1996) have measured the NLXG number density to a limiting flux of
$2\times 10^{-15} $erg$\,$s$^{-1}$cm$^{-2}$. Both of these
measurements are plotted on Figure 1, in good agreement with the
predictions of the Griffiths et al. model, but suggesting tighter
constraints on the upper and lower bounds.

To constrain the NLXG contribution further we vary the faint
end slope of the luminosity function to obtain an X-ray number-flux
relation giving a best fit through the two fainter measurements of
Almaini et al. (1997) and Hasinger (1996) (see Figure 1).  The revised
$1 \sigma$ error bounds are displayed as the shaded area in Figure
1. By integration of the predicted distribution to $10^{-20}
$erg$\,$s$^{-1}$cm$^{-2}$ these error bounds correspond to $28-55$ per
cent of the XRB at 1keV. The revised upper limits now $\em do\, not$
over-predict the XRB at 1keV.  In the next section we convert these
predictions into optical number counts.

\section{The predicted optical number-magnitude relation}

To convert the X-ray number counts into the optical band we use the
X-ray to optical ratio of NLXGs from the Deep $\em ROSAT$ Survey of
Shanks et al. (1997):

\begin{equation}
f_{\nu}(2$keV$)/f_{\nu}(4000{\rm{\AA}})=10^{4.7\pm0.8}.
\end{equation}

where the error quoted represents the dispersion in X-ray to optical
ratios rather than the error on the mean. We adopt an optical spectral
index of $\alpha_o=1$, which assumes that these are predominantly
late-type/Irr galaxies similar to those of Tresse et
al. (1996). Locally observed Seyfert 2 galaxies have similar
optical/UV spectra (Kinney et al. 1991, Heckman et al. 1995).

Since the galaxies in the HDF are believed to be mostly at high
redshift ($z > 1$) we choose to convert the X-ray number counts into
optical predictions for the $I$ band, corresponding to rest-frame
wavelengths in the optical/ultra-violet region. To include the effect
of this dispersion in the X-ray to optical spectral index we assume a
Gaussian distribution in the log of the flux ratio given above with a
dispersion $\sigma=0.8$ and distribute the optical counts
accordingly. The resulting number-magnitude relations are shown in
Figure 2, with $1 \sigma$ error bounds given by the shaded region.
Thus in the F814 HDF image, with an approximate magnitude limit of
$I=28$, we predict $\sim 10^3$ QSOs deg$^{-2}$ but a considerably
larger number of NLXGs ($\sim 10^{5.3\pm0.2}$ deg$^{-2}$). With a more
conservative lower limit of $\sim 10^{40}$erg s$^{-1}$ (rather than
$10^{39}$erg s$^{-1}$) in the X-ray luminosity function, the predicted
counts are reduced so that the centre of the shaded region will lie on
the dashed line in Figure 2.

\begin{figure}
\centering \centerline{\epsfxsize=12truecm
\figinsert{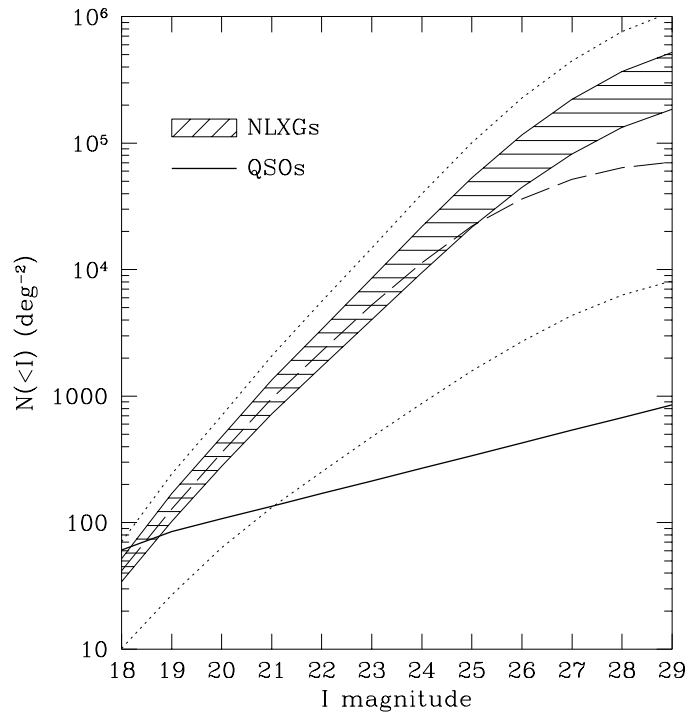}{0.0pt}}
\caption{Showing the predicted I-band number-magnitude relation for
QSOs and NLXGs. The dotted lines show the $1\sigma$ errors arising
from the faint end slope of the NLXG luminosity function of Griffiths
et al. (1996).  The shaded area shows the tighter predictions which
arise if we impose the measurements by Almaini et al. (1997) and
Hasinger (1996) (see Figure 1).  The dashed line shows the effect of
increasing the lower limit in the luminosity function from
$10^{39}$erg s$^{-1}$ to $10^{40}$erg s$^{-1}$.  }
\end{figure}

\section{Summary and conclusions}

There is now very strong evidence that a population of X-ray luminous
narrow emission-line galaxies could account for the origin of the
cosmic X-ray background.  We use the latest measurements of the X-ray
luminosity function for NLXGs, combined with their number density at
the faintest X-ray fluxes, to predict the number of objects expected
in deep optical images such as the HDF. Using measurements of the  X-ray
to optical luminosity ratio, we estimate that this new class of AGN
could outnumber broad-line QSOs by two orders of magnitude at a limit
of $I\sim28$, accounting for $\sim 10$ per cent of the faint galaxies
in the HDF. We note that if a significant fraction of these galaxies
have since merged then the percentage of present day quiescent
galaxies which have at one time hosted an X-ray luminous AGN could be
much higher.

With current technology we cannot expect to identify the faintest HDF
objects by spectroscopy, but even by $I=22$ we expect $\sim10^3$ NLXGs
deg$^{-2}$, i.e. a few percent of all galaxies at this magnitude.  The
brighter X-ray luminous NLXGs identified so far are formally
classified as a mixture of starburst and Seyfert 2 galaxies on the
basis of standard emission-line ratios (Boyle et al. 1995a).  In terms
of their optical magnitudes, X-ray properties and redshift
distributions, however, the two populations cannot be
distinguished. As noted by Boyle et al. (1995a), it seems likely that
one underlying physical mechanism is at work.  Faint optical redshift
surveys (e.g. Tresse et al. 1996) have shown that a large fraction
($\sim 50$ per cent) of all faint field galaxies show only emission
line features. Differentiating potentially X-ray luminous NLXGs from
the field population may therefore be difficult without supporting
high resolution X-ray observations. Further work is clearly required
to establish a reliable diagnostic method for identifying this `new'
class of active galaxies.

Previous predictions of faint X-ray and optical source populations
have been made by Treyer \& Silk (1993) on the assumption of a
population of evolving starbursting dwarf galaxies powered by short
lived, massive stars. Such models have difficulty in explaining the
XRB however, since ASCA observations of starburst galaxies have
revealed soft X-ray spectra which cannot contribute significantly to
the hard XRB (eg. Dela-Ceca et al 1996).

Near infra-red spectroscopy of local NLXGs may provide a conclusive
test of the obscured AGN hypothesis by allowing us to see through any
obscuring dust to detect broad emission lines in moderately obscured
AGN. If the nuclei are heavily obscured then the near infra-red
emission may also be obliterated, but in this scenario it is difficult
to produce the soft X-ray flux we observe with $\em ROSAT$ without a
very large ($\sim 10$ per cent) scattered component. The advection
dominated accretion flow model of Di Matteo \& Fabian (1997) should
also be tested by searching for the inverted radio spectra predicted
at short radio wavelengths.

We suggest that future workers on the HDF should be aware that $\sim
10$ per cent of the objects are potentially X-ray active and may
contain AGN. Obscured QSOs and/or advection dominated accretion models
are currently the most likely explanations for these unusual
galaxies. If the latter model is correct, we predict that compact
radio sources may be coincident with the more active NLXGs in the
HDF. Given the subarcsecond resolution of the forthcoming AXAF X-ray
satellite, a deep ($\sim1$Ms) observation of the HDF will be able to
identify these X-ray sources to a flux limit of $\sim3\times 10^{-17}
$erg$\,$s$^{-1}$cm$^{-2}$ ($0.5-2\,$keV)
 and hence test the predicted number-flux
relation displayed in Figure 1. Making use of the follow up work
already achieved on HDF, we anticipate that such an observation will
rapidly establish the true nature of the XRB.

\section*{ACKNOWLEDGMENTS}

ACF thanks the Royal Society for support. OA was funded by a PPARC
postdoctoral fellowship.

\end{document}

%% file: epsf_mn.tex
\newread\epsffilein    
\newif\ifepsffileok    
\newif\ifepsfbbfound   
\newif\ifepsfverbose   
\newdimen\epsfxsize    
\newdimen\epsfysize    
\newdimen\epsftsize    
\newdimen\epsfrsize    
\newdimen\epsftmp      
\newdimen\pspoints     
\def\epsfbox#1{\global\def\epsfllx{72}\global\def\epsflly{72}%
   \global\def\epsfurx{540}\global\def\epsfury{720}%
   \def\lbracket{[}\def\testit{#1}\ifx\testit\lbracket
   \let\next=\epsfgetlitbb\else\let\next=\epsfnormal\fi\next{#1}}%
\def\epsfgetlitbb#1#2 #3 #4 #5]#6{\epsfgrab #2 #3 #4 #5 .\\%
   \epsfsetgraph{#6}}%
\def\epsfnormal#1{\epsfgetbb{#1}\epsfsetgraph{#1}}%
\def\epsfgetbb#1{%
\openin\epsffilein=#1
\ifeof\epsffilein\errmessage{I couldn't open #1, will ignore it}\else
   {\epsffileoktrue \chardef\other=12
    \def\do##1{\catcode`##1=\other}\dospecials \catcode`\ =10
    \loop
       \read\epsffilein to \epsffileline
       \ifeof\epsffilein\epsffileokfalse\else
          \expandafter\epsfaux\epsffileline:. \\%
       \fi
   \ifepsffileok\repeat
   \ifepsfbbfound\else
    \ifepsfverbose\message{No bounding box comment in #1; using defaults}\fi\fi
   }\closein\epsffilein\fi}%
\def\epsfsetgraph#1{%
   \epsfrsize=\epsfury\pspoints
   \advance\epsfrsize by-\epsflly\pspoints
   \epsftsize=\epsfurx\pspoints
   \advance\epsftsize by-\epsfllx\pspoints
   \epsfxsize\epsfsize\epsftsize\epsfrsize
   \ifnum\epsfxsize=0 \ifnum\epsfysize=0
      \epsfxsize=\epsftsize \epsfysize=\epsfrsize
     \else\epsftmp=\epsftsize \divide\epsftmp\epsfrsize
       \epsfxsize=\epsfysize \multiply\epsfxsize\epsftmp
       \multiply\epsftmp\epsfrsize \advance\epsftsize-\epsftmp
       \epsftmp=\epsfysize
       \loop \advance\epsftsize\epsftsize \divide\epsftmp 2
       \ifnum\epsftmp>0
          \ifnum\epsftsize<\epsfrsize\else
             \advance\epsftsize-\epsfrsize \advance\epsfxsize\epsftmp \fi
       \repeat
     \fi
   \else\epsftmp=\epsfrsize \divide\epsftmp\epsftsize
     \epsfysize=\epsfxsize \multiply\epsfysize\epsftmp   
     \multiply\epsftmp\epsftsize \advance\epsfrsize-\epsftmp
     \epsftmp=\epsfxsize
     \loop \advance\epsfrsize\epsfrsize \divide\epsftmp 2
     \ifnum\epsftmp>0
        \ifnum\epsfrsize<\epsftsize\else
           \advance\epsfrsize-\epsftsize \advance\epsfysize\epsftmp \fi
     \repeat     
   \fi
   \ifepsfverbose\message{#1: width=\the\epsfxsize, height=\the\epsfysize}\fi
   \epsftmp=10\epsfxsize \divide\epsftmp\pspoints
   \newcount\figskipcount
      \message{#1 \the\epsfysize  }
   \vbox to\epsfysize{\vfil\hbox to\epsfxsize{%
      \includegraphics{#1}%
      \hfil}}%
\epsfxsize=0pt\epsfysize=0pt}%

{\catcode`\%=12 \global\let\epsfpercent=
\long\def\epsfaux#1#2:#3\\{\ifx#1\epsfpercent
   \def\testit{#2}\ifx\testit\epsfbblit
      \epsfgrab #3 . . . \\%
      \epsffileokfalse
      \global\epsfbbfoundtrue
   \fi\else\ifx#1\par\else\epsffileokfalse\fi\fi}%
\def\epsfgrab #1 #2 #3 #4 #5\\{%
   \global\def\epsfllx{#1}\ifx\epsfllx\empty
      \epsfgrab #2 #3 #4 #5 .\\\else
   \global\def\epsflly{#2}%
   \global\def\epsfurx{#3}\global\def\epsfury{#4}\fi}%
\def\epsfsize#1#2{\epsfxsize}